\documentclass[twocolumn,3p,times]{elsarticle}

\usepackage{amssymb}
\usepackage{graphics}
\usepackage{graphicx}
\usepackage{txfonts}
\usepackage{epsfig}
\usepackage{bm}   
\usepackage{dcolumn}
\usepackage{rotate}
\usepackage{amsfonts}
\usepackage{longtable}
\usepackage{latexsym,rotating,wrapfig}
\usepackage{footmisc,booktabs,amssymb,longtable}

\usepackage{xcolor,colortbl}

\usepackage{multirow}
\definecolor{Gray}{gray}{0.85}
\definecolor{LightCyan}{rgb}{0.88,1,1}

\usepackage[scaled=0.92]{helvet}
\usepackage[T1]{fontenc}
\usepackage{textcomp}
\usepackage{braket}
\usepackage{textgreek}
\usepackage{rotate}
\usepackage{tabularx}
\usepackage[referable]{threeparttablex}
\usepackage{epigraph}
\begin{document}

\begin{frontmatter}

\title{Subshell gaps and onsets of collectivity \\ from proton and neutron pairing gap correlations}

\author{Jos\'e Nicol\'as Orce}
\ead{jnorce@uwc.ac.za; nico.orce@cern.ch; http://nuclear.uwc.ac.za; https://github.com/UWCNuclear}
\address{Department of Physics \& Astronomy, University of the Western Cape, P/B X17, Bellville 7535, South Africa}
\address{National Institute for Theoretical and Computational Sciences (NITheCS), South Africa}

\begin{abstract}

Throughout the nuclear chart, particle-hole correlations give rise to giant resonances and, together with the proton-neutron interaction, deformation and rotational bands.
In order to shed light on many-body correlations in open-shell nuclei, I explore macroscopic properties that could manifest from the collective behaviour of protons and neutrons.
Intuitively, the correlation of proton and neutron Cooper pairs can be inferred from the respective pairing gaps, that can precisely be extracted  from the AME 2020 atomic
mass evaluation through odd-even atomic mass differences. This work shows that the combination of large and close-lying proton and neutron pairing gaps is
sensitive to onsets of collectivity and subshell gaps in superfluid nuclei, away from major shell closures.
Trends of reduced transition probabilities or B(E2) values --- which describe the collective overlap between the wave functions of initial and final nuclear states ---
are revealed in overall agreement with data.
Specially interesting is the peak of collectivity in the tin isotopes at $^{110}$Sn, instead of at midshell, as expected by large-scale shell-model calculations;
a situation that has astounded the nuclear physics community for quite some time.

\end{abstract}


\end{frontmatter}



Despite the lack of a definite non-local two-nucleon ({\sc NN}) interaction,
a new and exciting era has started for microscopic nuclear structure physics with
advancements arising from chiral effective field theory~\cite{machleidt2023ab,tichai2024towards}, renormalization methods~\cite{brueckner1955approximate,brueckner1955two,bethe1963reference,hergert2013medium}
and high-performance supercomputing~\cite{schneider2022exascale}.
Nonetheless, calculations of nuclear structure properties away from light nuclei and shell closures remain challenging, more often impossible,
because of the impracticality for a microscopic description of open-shell, heavy nuclei~\cite{greiner1996nuclear,tichai2020many,coraggio2021future}.
Renormalization   techniques may be missing some physics while integrating out the unsettling infinities and smoothing down high-order terms such as matrix elements
arising from two-pion exchange. Modern supercomputing technologies and numerical algorithms cannot diagonalize the Schr\"odinger equation with the large matrix dimensions (> 10$^{12}$)
involved in the many-body correlations of open-shell nuclei.
The challenge is reaching numerical convergence of physical observables in the infinite-dimensional Hilbert basis space.
This is probably the reason
for the underestimation of collective properties such as electric quadrupole ({\sc E2}) reduced transition probabilities --- or B(E2) values --- and spectroscopic quadrupole moments~\cite{henderson2018testing}.


We are urged to explore alternative ideas that may uncover the collective behaviour of protons and neutrons in open-shell nuclei.
Inasmuch the microscopic behaviour of individual atoms and molecules  manifest through macroscopic average properties such as mass, temperature, electric-quadrupole or magnetic-dipole moments,
such properties can also be determined in nuclear systems, 
and provide information on how the nucleus behaves collectively~\cite{pritychenko2016tables,qmoments,deformation,harakeh2001giant,bortignon2019giant}.

In the current study, I explore weak-collective systems characterized by small B(E2) values, where the quadrupole-quadrupole interaction does not give rise to large deformations.
In particular, I investigate a key property of open-shell nuclides in the Ni, Cd, Sn and Te isotopes; namely, nuclear superfluidity.
Remarkably, unlike the smooth behaviour of excitation energies and neutron separation energies within an isotopic chain, correlations between  proton and neutron pairing gaps show high sensitivity to onsets of nuclear collectivity and subshell gaps in open-shell nuclei. 
This study expands on previous work~\cite{orce2014strong} by considering two major upgrades on data evaluation~\cite{pritychenko2016tables,wang2021ame}.
Namely, the 2016 tables of B(E2) values from the first excited states in even–even nuclei~\cite{pritychenko2016tables} and the 2020 Atomic Mass Evaluation ~\cite{wang2021ame}, which substantially improve previous ones~\cite{audi2012ame2012,wang2012ame2012,raman} in accuracy and range.

\begin{figure}[!ht]
\begin{center}
\includegraphics[width=7cm,height=5.5cm,angle=-0]{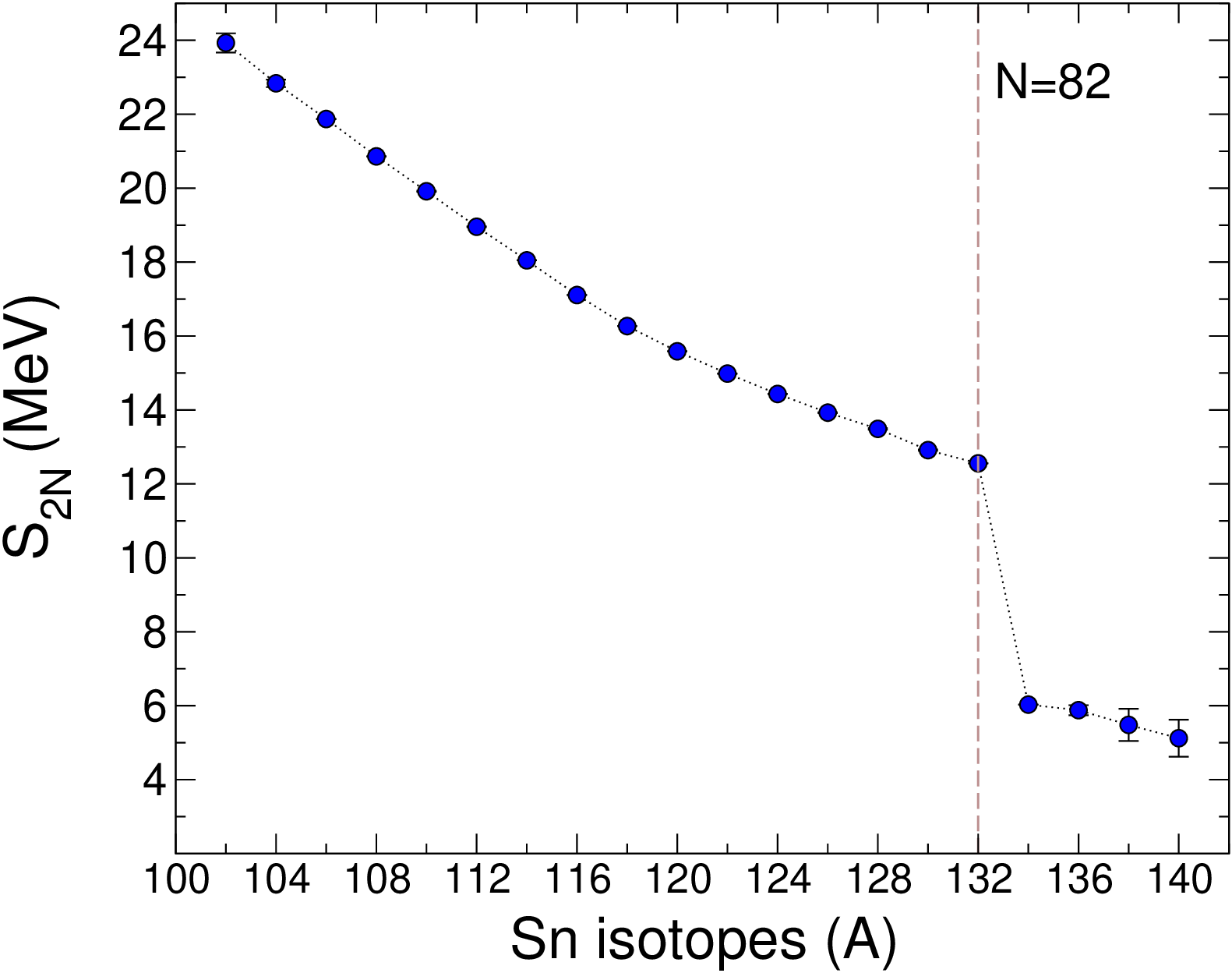}
\caption{Two-neutron separation energies ($S_{2N}$) for the $^{102-140}$Sn even-even isotopes extracted from the 2020 atomic mass evaluation~\cite{wang2021ame},
presenting a smooth trend from $^{102-130}$Sn and a dramatic drop at $N=82$ (dashed vertical line). The dotted line is shown for visual purposes.}
\label{fig:s2n}
\end{center}
\end{figure}

Average nuclear properties such as charge radii and quadrupole deformations are closely related, through a density function, 
to  ground-state masses~\cite{deformation,thomasfermi,hfbcs,hfbcs2}. 
The macroscopic finite-range droplet with Lipkin-Nogami
pairing~\cite{deformation} and Thomas-Fermi~\cite{thomasfermi} models with microscopic corrections can  calculate reasonably well
quadrupole deformations with  parameters fitted exclusively to  mass data.
Moreover, microscopic approaches based on Hartree Fock with BCS pairing  provides good results for charge radii and
quadrupole deformations~\cite{hfbcs}.
The deformation energy can be defined as the difference between the masses for the spherical configuration,
$M^{sph}$, and the mass calculated at the equilibrium, $M^{equil}$,
\begin{equation}
E_{def}=M^{sph}-M^{equil}\ >0.
\end{equation}
Through proton and neutron separation energies, atomic masses are sensitive indicators of nuclear structural changes within an isotopic chain,
either by pointing at shell gaps~\cite{hakala2008evolution} or an onset of deformation~\cite{hager}.
Nevertheless, any structural changes remain inconspicuous for open-shell nuclei, as shown in Fig.~\ref{fig:s2n} for the tin isotopes
by the smooth behaviour of the two-neutron separation energies, $S_{2N}$, below $^{132}$Sn.


\begin{figure*}[!ht]
\begin{center}
\includegraphics[width=7.3cm,height=6.5cm,angle=0]{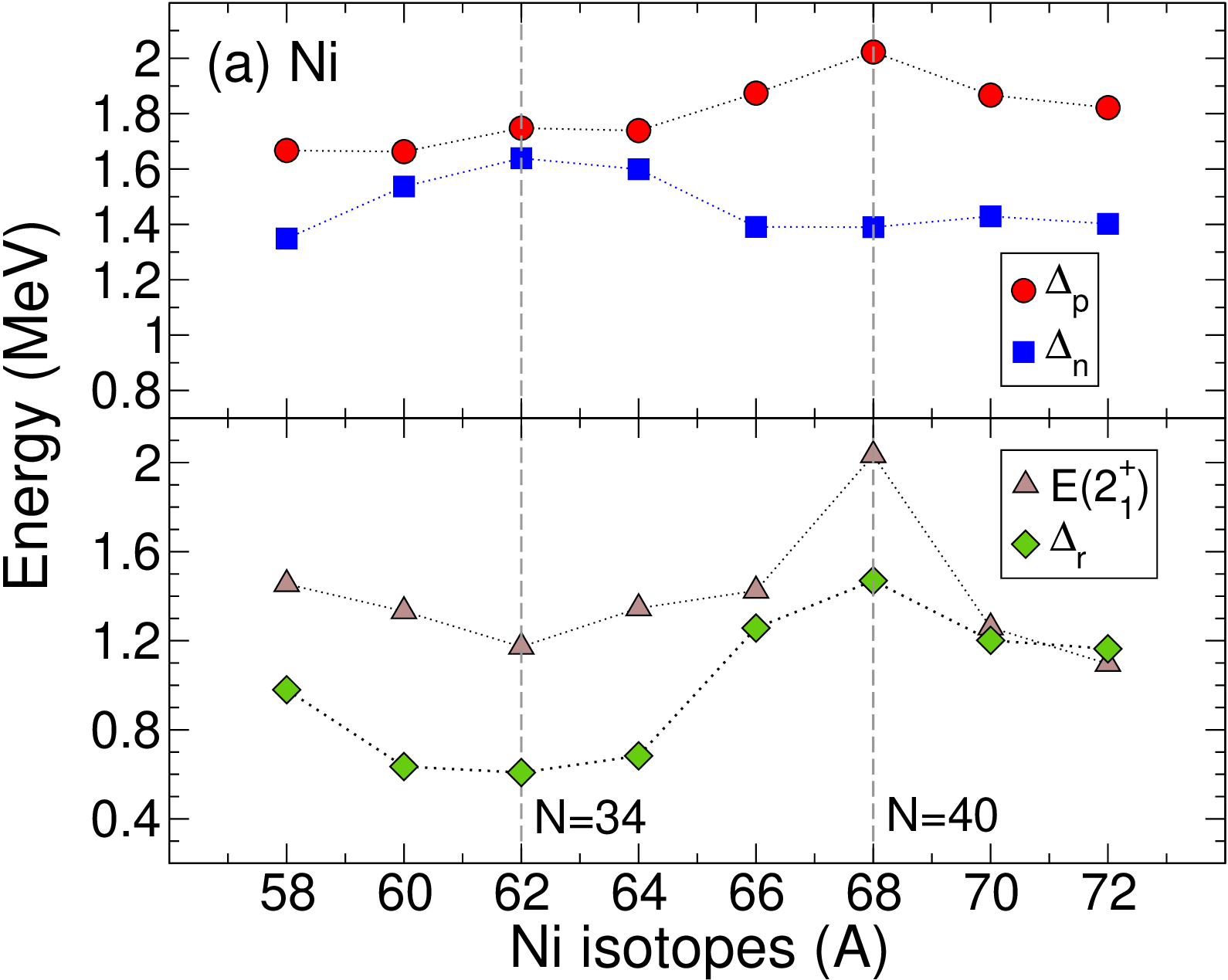}
\hspace{0.4cm}
\includegraphics[width=7.cm,height=6.5cm,angle=0]{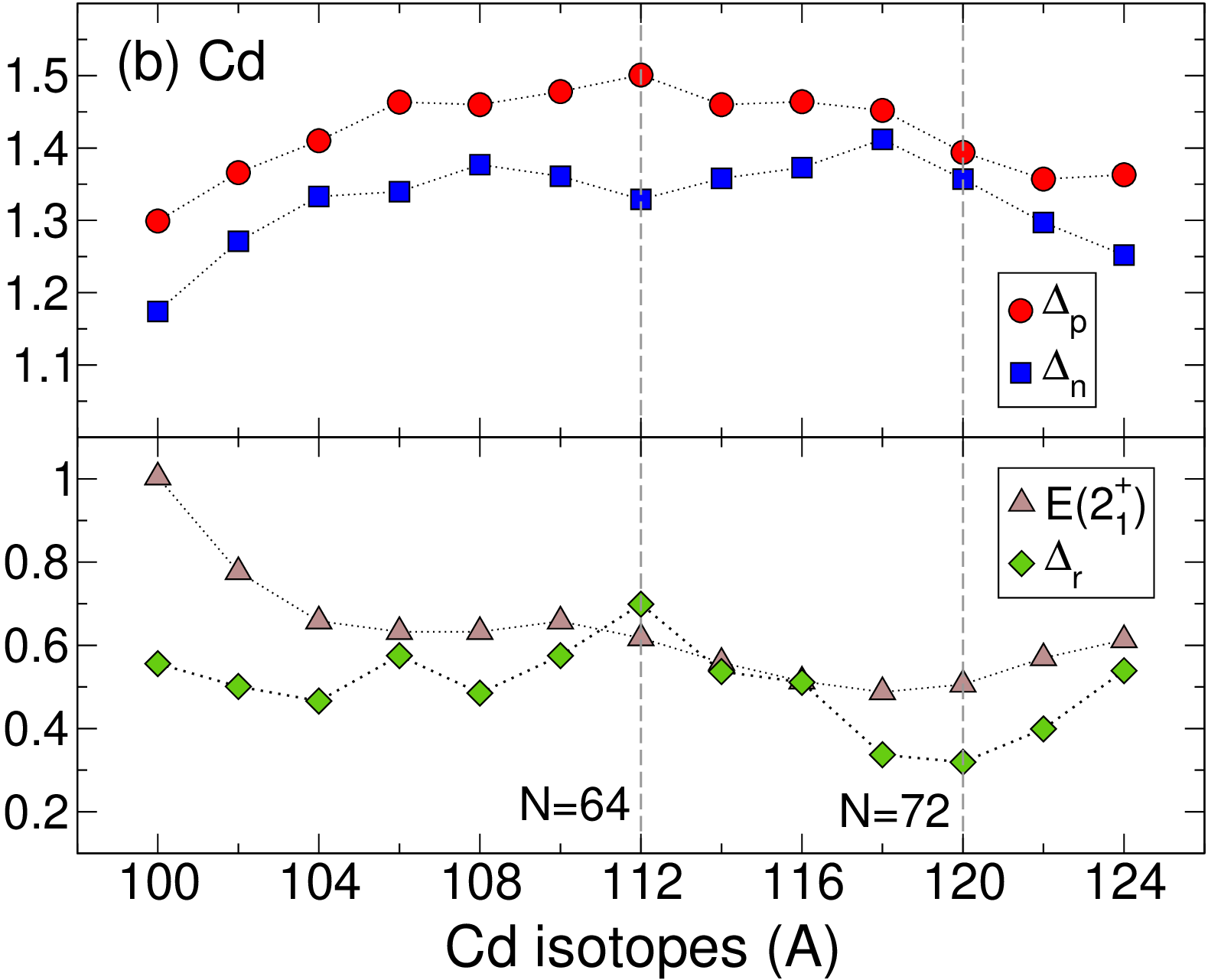}\\
\vspace{0.7cm}
\includegraphics[width=7.3cm,height=6.5cm,angle=0]{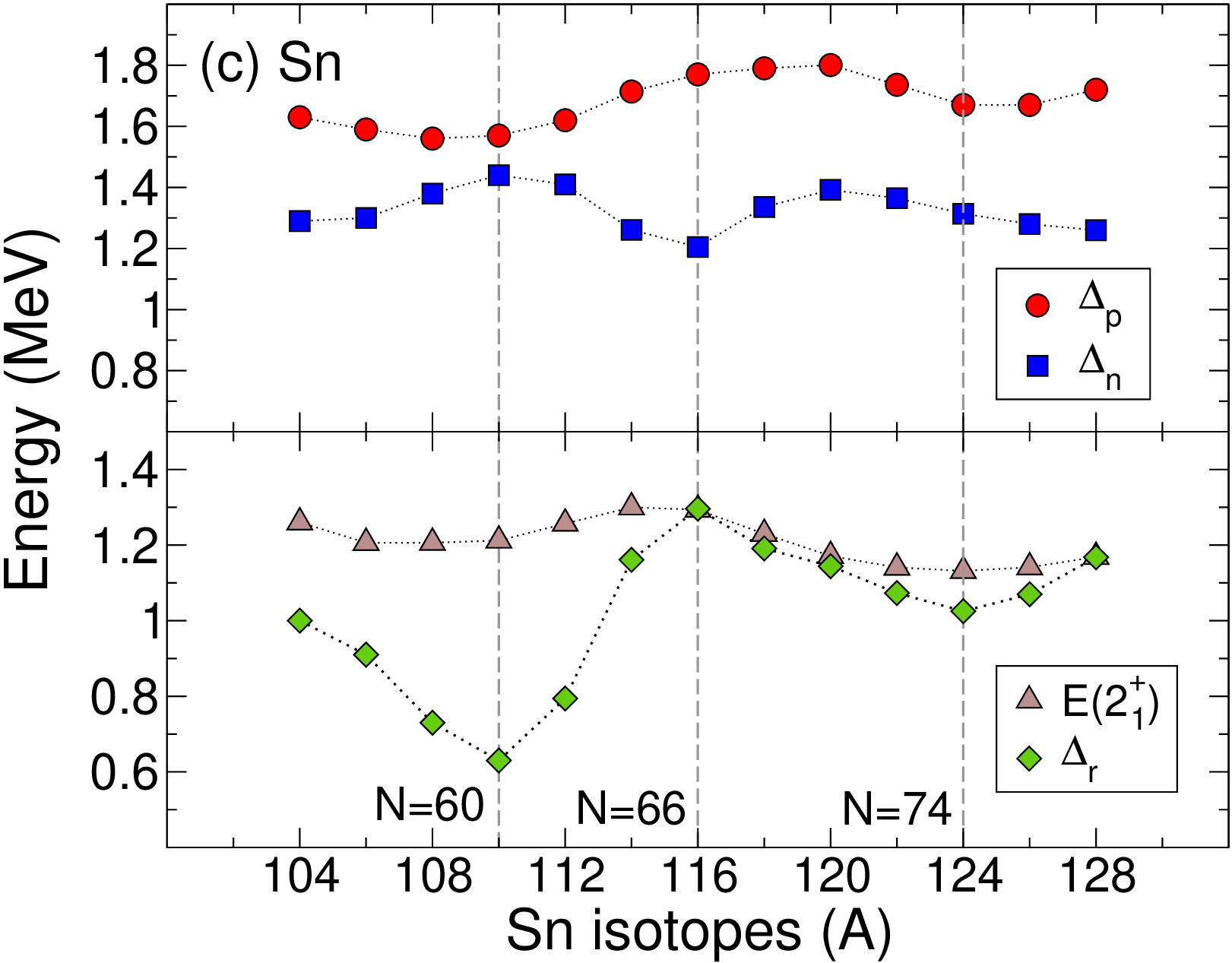}
\hspace{0.4cm}
\includegraphics[width=7.cm,height=6.5cm,angle=-0]{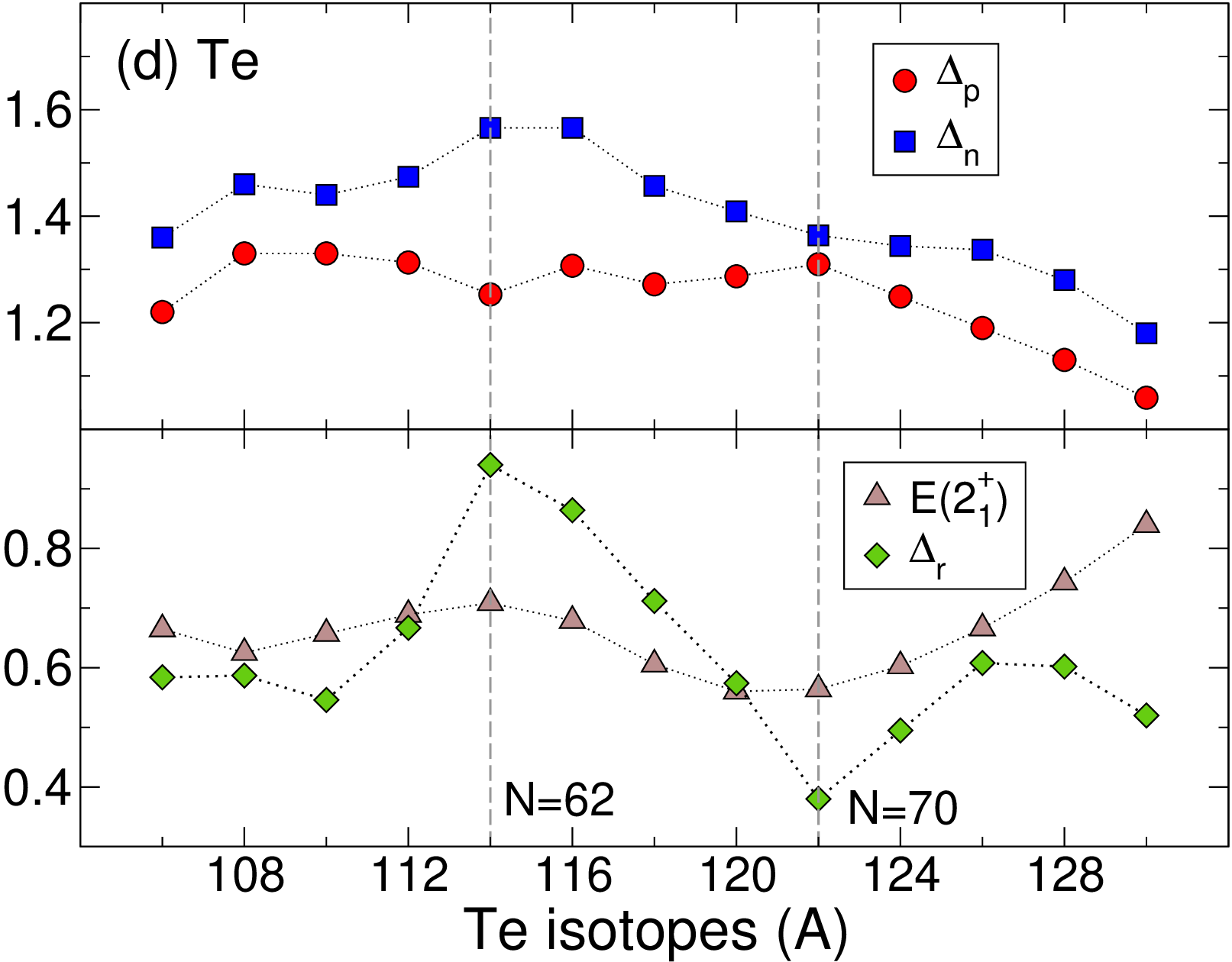}
\caption{The top panels show proton (circles) and neutron (squares) pairing gaps from the {\sc AME} 2020 Atomic Mass Evaluation~\cite{wang2021ame},
whereas the bottom panels show a comparison of E(2$^+_1$) (triangles up) and $\Delta_r$ (diamonds) energies for the even-mass (a) Ni, (b) Cd, (c) Sn and (d) Te isotopes.}
\label{fig:3}
\end{center}
\end{figure*}

Within the BCS pairing model~\cite{bcs}, the first excitation in deformed nuclei --- generally $J^{\pi}=2^+_1$ --- arises by breaking one Cooper pair,
and is interpreted as a two quasiparticle which lies at least at the pairing energy, $2\Delta$.
%
The magnitude of the neutron, $\Delta_n$, and proton, $\Delta_p$, pairing gaps can be determined
from experimental odd-even mass differences derived from the
Taylor expansion of the nuclear mass in nucleon-number
differences~\cite{deformation}.
These prescriptions assume that pairing is the only non-smooth
contribution to nuclear masses. For  weakly-deformed even-even nuclei away from neutron shell closures,
we extract $\Delta_n$ and $\Delta_p$ from the symmetric five-point mass difference~\cite{deformation,bender},
\begin{eqnarray}
\Delta_n &=& -\frac{1}{8} [M(Z,N+2)-4M(Z,N+1)+6M(Z,N) \nonumber \\
&-& 4M(Z,N-1)+M(Z,N-2) ]\\
\Delta_p &=& -\frac{1}{8} [M(Z+2,N)-4M(Z+1,N)+6M(Z,N) \nonumber \\
&-& 4M(Z-1,N)+M(Z-2,N) ].
\end{eqnarray}
Here, we make the strong assumption of a valid $\Delta_p$ in the region of study,
although the kink of binding energies at shell closures would a priori not allow a Taylor expansion.
This assumption is somewhat supported by the $^{56}$Ni~\cite{kenn,otsuka,honma0} and $^{100}$Sn~\cite{sn_msu,bader2013quadrupole}
soft cores and Refs.~\cite{moller1992nuclear,bender2000pairing}.
Doubly magic nuclei  are not included in the pairing-gap systematics
since in these cases both magic-number and Wigner cusps span
singularities in the mass surface~\cite{madlandnix}.
The top panels of Fig.~\ref{fig:3} show $\Delta_n$ and $\Delta_p$ values in the even-mass (a) Ni, (b) Cd, (c) Sn and (d) Te isotopes.
Except for the  Te isotopes, $\Delta_p$ lies lower than the corresponding $\Delta_n$ values.

For nuclei where proton and neutron Fermi surfaces lie close to each other, one can assume that the interplay of both proton and neutron pairing gaps may contribute
to the collective origin of the 2$^+_1$ state. Accordingly, we introduce the {\it relative neutron-proton pairing gap}, $\Delta_r$, given by,
\begin{equation}
\Delta_r^2 := ~\mid (\Delta_p - \Delta_n) (\Delta_p + \Delta_n) \mid ~= ~\mid \Delta_p^2 - \Delta_n^2 \mid \label{eq:deltar},
\end{equation}
\noindent  where the first term ($\Delta_p - \Delta_n$) is the \emph{resonant factor}, which accounts for the
proximity of proton and neutron pairing-gaps. That is, the smaller ($\Delta_p - \Delta_n$), the larger the overlap between
proton and neutron pairing fields and the more collective is the nuclear system. The second term  ($\Delta_p + \Delta_n$) is the \emph{energy factor}, and accounts for the energy that can be provided to
the nuclear system before breaking Cooper pairs, i.e., the larger the ($\Delta_p + \Delta_n$), the less collective is the system.

The bottom panels of Fig.~\ref{fig:3} show the trends of $\Delta_r$ values and 2$^+_1$ excitation energies,
which follow similar, but not identical trends. Sharper energy differences are observed for $\Delta_r$ values,
with more sensitivity to subshell closures characterized by energy gaps at $N=40$, $N=64$, $N=66$ and $N=62$ for the Ni, Cd, Sn and Te isotopes, respectively.
Additionally, $\Delta_r$ values are clearly more sensitive to onsets of deformation, characterized by the proximity of $\Delta_p$ and $\Delta_n$ values,
at $N=34$, $N=72$, $N=60,74$ and $N=70$ in the Ni, Cd, Sn and Te isotopes, respectively. These neutron numbers correspond to peaks of collectivity, as observed by the trend of experimental $B(E2; 0_1^+ \rightarrow 2^+_1)$ values~\cite{pritychenko2016tables,maheshwari2016asymmetric,maheshwari2019evolution}.
Such trends of $B(E2; 0^+_1 \rightarrow 2^+_1)$ values in isotopic chains have been calculated using Grodzins' formula and variations of it~\cite{grodzins,raman,pritychenko2017revisiting},
and shown to be inversely proportional to $E(2^+_1)$. Here, we instead propose the use of $\Delta_r$ values, which clearly show a sharper distinction than excitation energies
for a particular isotopic chain. One can simply infer a similar collectivity for
$^{60,62,64}$Ni with $\Delta_r\approx 0.6$ MeV, and sharp minima at $^{118-120}$Cd, $^{110}$Sn and $^{122}$Te with $\Delta_r$ values of 0.32-0.34, 0.63 and 0.38 MeV, respectively.
This simple model fails for the neutron-rich Te isotopes, as the $N=82$ shell closure approaches and nuclei stop behaving like a superfluid.

The foundations for this remarkable behaviour predicted by pairing gaps are laid in my previous work~\cite{orce2014strong}, which has recently been supported by more precise calculations
using generalized seniority and large-scale shell-model calculations~\cite{maheshwari2016asymmetric,maheshwari2019evolution,zuker2021quadrupole}.
The converging idea lies in the Lipkin-Nogami pairing formalism~\cite{lipkin1960collective,nogami1964improved}, which allows the determination of proton pairing gaps at magic numbers~\cite{suhonen2007nucleons} and where $B(E2; 0_1^+ \rightarrow 2^+_1)$ values can be inferred as inversely proportional to the pair degeneracy
$2\Omega=(2j+1)$~\cite{maheshwari2016odd} and the pairing strength~\cite{zuker2021quadrupole}.\\

Finally, this interpretation considers pairing and quadrupole correlations but lacks the major source of deformation arising from the interaction of pairs of valence
$N_p$  protons and $N_n$  neutrons~\cite{hamamoto1965empirical,nair1977neutron,federman1977towards,casten1985npnn,heyde2011shape,bonatsos2015proton} --- namely, the product $N_pN_n$ ---
which can certainly influence these weakly-collective nuclides. The interaction of a pair of neutrons and protons can be associated with $\alpha$-type
correlations~\cite{lanealphawidhts,flowers1963charge}, which are also needed to describe $\alpha$ particle reduced widths~\cite{mang1960calculation,sandulescu1962reduced,sandulescu2012four,baran2016proton}.
In order to investigate the influence of $\alpha$ correlations is instructive to quantify the role of $\Delta_r$ in the collectivity of the 2$^+_1$ states.
This was partially done in my previous work~\cite{orce2014strong} for the Ni and Sn isotopes, where $B(E2; 0_1^+ \rightarrow 2^+_1) \propto \frac{1}{2\Delta_r}$,
showing agreement for the Ni isotopes, but deviations in the tin isotopes where $\alpha$ correlations are known to be more relevant~\cite{sandulescu2012four,baran2016proton,delion2009universal,delion2010investigations}.
More advanced large-scale-shell model calculations can nicely reproduce the  $B(E2; 0_1^+ \rightarrow 2^+_1)$ trend in the tin isotopes by truncating the model space
after fixing the orbits involved in the configuration mixing~\cite{maheshwari2016asymmetric}.
The relevant inclusion of $\alpha$-type correlations deserves further exploration that will be presented in a separate manuscript. \\

The author had the privilege to meet the late Harry Lipkin at Argonne National Laboratory in Chicago and
discuss his model on collective motion in many-body systems. He also acknowledges the single most prolific nuclear data evaluator, the late Balraj Singh,  for
physics discussions regarding the trend of B(E2) values and providing relevant information.

\bibliographystyle{elsarticle-num}

\bibliography{adndt}

\begin{thebibliography}{10}
\expandafter\ifx\csname url\endcsname\relax
  \def\url#1{\texttt{#1}}\fi
\expandafter\ifx\csname urlprefix\endcsname\relax\def\urlprefix{URL }\fi
\expandafter\ifx\csname href\endcsname\relax
  \def\href#1#2{#2} \def\path#1{#1}\fi

\bibitem{machleidt2023ab}
R.~Machleidt, What is ab initio?, Few-Body Systems 64~(4) (2023) 77.

\bibitem{tichai2024towards}
A.~Tichai, P.~Demol, T.~Duguet, Towards heavy-mass nuclear structure:
  Open-shell ca, ni and sn isotopes from bogoliubov coupled-cluster theory,
  Physics Letters B 1001 (2024).

\bibitem{brueckner1955approximate}
K.~Brueckner, C.~Levinson, Approximate reduction of the many-body problem for
  strongly interacting particles to a problem of self-consistent fields,
  Physical Review 97~(5) (1955) 1344.

\bibitem{brueckner1955two}
K.~Brueckner, Two-body forces and nuclear saturation. iii. details of the
  structure of the nucleus, Physical Review 97~(5) (1955) 1353.

\bibitem{bethe1963reference}
H.-A. Bethe, B.~Brandow, A.~Petschek, Reference spectrum method for nuclear
  matter, Physical Review 129~(1) (1963) 225.

\bibitem{hergert2013medium}
H.~Hergert, S.~Bogner, S.~Binder, A.~Calci, J.~Langhammer, R.~Roth, A.~Schwenk,
  In-medium similarity renormalization group with chiral two-plus three-nucleon
  interactions, Physical Review C—Nuclear Physics 87~(3) (2013) 034307.

\bibitem{schneider2022exascale}
D.~Schneider, The exascale era is upon us: The frontier supercomputer may be
  the first to reach 1,000,000,000,000,000,000 operations per second, IEEE
  spectrum 59~(1) (2022) 34--35.

\bibitem{greiner1996nuclear}
W.~Greiner, J.~A. Maruhn, et~al., Nuclear models, Vol. 261, Springer, 1996.

\bibitem{tichai2020many}
A.~Tichai, R.~Roth, T.~Duguet, Many-body perturbation theories for finite
  nuclei, Frontiers in Physics 8 (2020) 164.

\bibitem{coraggio2021future}
L.~Coraggio, S.~Pastore, C.~Barbieri, The future of nuclear structure:
  Challenges and opportunities in the microscopic description of nuclei,
  Frontiers in Physics 8 (2021) 626976.

\bibitem{henderson2018testing}
J.~Henderson, G.~Hackman, P.~Ruotsalainen, S.~Stroberg, K.~Launey, J.~Holt,
  F.~Ali, N.~Bernier, M.~Bentley, M.~Bowry, et~al., Testing microscopically
  derived descriptions of nuclear collectivity: Coulomb excitation of 22mg,
  Physics Letters B 782 (2018) 468--473.

\bibitem{pritychenko2016tables}
B.~Pritychenko, M.~Birch, B.~Singh, M.~Horoi, Tables of e2 transition
  probabilities from the first 2+ states in even--even nuclei, At. Data Nucl.
  Data Tables 107 (2016) 1--139.

\bibitem{qmoments}
N.~J. Stone, Table of nuclear magnetic dipole and electric quadrupole moments,
  Atomic Data Nuclear Data Tables 111-112 (2016) 1--28.

\bibitem{deformation}
P.~M\"oller, J.~R. Nix, W.~D. Myers, W.~J. Swiatecki, Nuclear ground-state
  masses and deformations, At. Data Nucl. Data Tables 59 (1995) 185.

\bibitem{harakeh2001giant}
M.~N. Harakeh, A.~Woude, Giant Resonances: fundamental high-frequency modes of
  nuclear excitation, Vol.~24, Oxford Studies in Nuclear Phys, 2001.

\bibitem{bortignon2019giant}
P.~F. Bortignon, A.~Bracco, R.~A. Broglia, Giant Resonances: Nuclear structure
  at finite temperature, CRC Press, 2019.

\bibitem{orce2014strong}
J.~N. Orce, A strong correlation between nuclear collectivity and atomic
  masses, Journal of Physics G 41~(5) (2014) 055113.

\bibitem{wang2021ame}
M.~Wang, W.~J. Huang, F.~G. Kondev, G.~Audi, S.~Naimi, The ame 2020 atomic mass
  evaluation (ii). tables, graphs and references, Chinese Physics C 45~(3)
  (2021) 030003.

\bibitem{audi2012ame2012}
G.~Audi, M.~Wang, A.~Wapstra, F.~G. Kondev, M.~MacCormick, X.~Xu, B.~Pfeiffer,
  The ame2012 atomic mass evaluation, Chinese physics C 36~(12) (2012) 1287.

\bibitem{wang2012ame2012}
M.~Wang, G.~Audi, A.~Wapstra, F.~Kondev, M.~MacCormick, X.~Xu, B.~Pfeiffer, The
  ame2012 atomic mass evaluation, Chinese physics C 36~(12) (2012) 1603.

\bibitem{raman}
S.~Raman, C.~W. Nestor~Jr, P.~Tikkanen, Transition probability from the ground
  to the first-excited 2+ state of even--even nuclides, Atomic Data and Nuclear
  Data Tables 78~(1) (2001) 1--128.

\bibitem{thomasfermi}
W.~D. Myers, W.~J. Swiatecki, Nuclear properties according to the thomas-fermi
  model, Nuclear Physics A 601~(2) (1996) 141--167.

\bibitem{hfbcs}
S.~Goriely, F.~Tondeur, J.~M. Pearson, A hartree--fock nuclear mass table,
  Atomic Data and Nuclear Data Tables 77~(2) (2001) 311--381.

\bibitem{hfbcs2}
S.~Goriely, M.~Samyn, J.~M. Pearson, Further explorations of
  skyrme-hartree-fock-bogoliubov mass formulas. vii. simultaneous fits to
  masses and fission barriers, Physical Review C 75~(6) (2007) 064312.

\bibitem{hakala2008evolution}
J.~Hakala, S.~Rahaman, V.-V. Elomaa, T.~Eronen, U.~Hager, A.~Jokinen,
  A.~Kankainen, I.~Moore, H.~Penttil{\"a}, S.~Rinta-Antila, et~al., Evolution
  of the n= 50 shell gap energy towards ni 78, Physical review letters 101~(5)
  (2008) 052502.

\bibitem{hager}
U.~Hager, T.~Eronen, J.~Hakala, A.~Jokinen, V.~S. Kolhinen, S.~Kopecky,
  I.~Moore, A.~Nieminen, M.~Oinonen, S.~Rinta-Antila, et~al., First precision
  mass measurements of refractory fission fragments, Physical Review Letters
  96~(4) (2006) 042504.

\bibitem{bcs}
L.~N. Cooper, J.~Bardeen, J.~R. Schrieffer, Theory of superconductivity,
  Physical Review 108 (1957) 1175.

\bibitem{bender}
M.~Bender, K.~Rutz, P.~G. Reinhard, J.~A. Maruhn, Pairing gaps from nuclear
  mean-fieldmo dels, The European Physical Journal A 8 (2000) 59--75.

\bibitem{kenn}
O.~Kenn, K.-H. Speidel, R.~Ernst, J.~Gerber, N.~Benczer-Koller, G.~Kumbartzki,
  P.~Maier-Komor, F.~Nowacki, Striking harmony between the nuclear shell model
  and new experimental g factors and {B}({E}2) values of even {N}i isotopes,
  Physical Review C 63~(2) (2000) 021302(R).

\bibitem{otsuka}
T.~Otsuka, M.~Honma, T.~Mizusaki, Structure of the {N= Z= 28} closed shell
  studied by monte carlo shell model calculation, Physical Review Letters
  81~(8) (1998) 1588.

\bibitem{honma0}
M.~Honma, T.~Otsuka, B.~A. Brown, T.~Mizusaki, New effective interaction for
  pf-shell nuclei and its implications for the stability of the {N= Z= 28}
  closed core, Physical Review C 69~(3) (2004) 034335.

\bibitem{sn_msu}
C.~Vaman, C.~Andreoiu, D.~Bazin, A.~Becerril, B.~A. Brown, C.~M. Campbell,
  A.~Chester, J.~M. Cook, D.~C. Dinca, A.~Gade, et~al., {Z= 50} shell gap near
  {S}n 100 from intermediate-energy coulomb excitations in even-mass {S}n
  106-112 isotopes, Physical Review Letters 99~(16) (2007) 162501.

\bibitem{bader2013quadrupole}
V.~M. Bader, A.~Gade, D.~Weisshaar, B.~A. Brown, T.~Baugher, D.~Bazin,
  J.~Berryman, A.~Ekstr{\"o}m, M.~Hjorth-Jensen, S.~R. Stroberg, et~al.,
  Quadrupole collectivity in neutron-deficient {S}n nuclei: 104{S}n and the
  role of proton excitations, Physical Review C 88~(5) (2013) 051301.

\bibitem{moller1992nuclear}
P.~M{\"o}ller, J.~R. Nix, Nuclear pairing models, Nuclear Physics A 536~(1)
  (1992) 20--60.

\bibitem{bender2000pairing}
M.~Bender, K.~Rutz, P.~G. Reinhard, J.~A. Maruhn, Pairing gaps from nuclear
  mean-fieldmo dels, The European Physical Journal A 8 (2000) 59--75.

\bibitem{madlandnix}
D.~G. Madland, J.~R. Nix, New model of the average neutron and proton pairing
  gaps, Nuclear Physics A 476~(1) (1988) 1--38.

\bibitem{maheshwari2016asymmetric}
B.~Maheshwari, A.~K. Jain, B.~Singh, Asymmetric behavior of the {B}({E}2; 0+→
  2+) values in 104-130{S}n and generalized seniority, Nuclear Physics A 952
  (2016) 62--69.

\bibitem{maheshwari2019evolution}
B.~Maheshwari, H.~A. Kassim, N.~Yusof, A.~K. Jain, Evolution of nuclear
  structure in and around z= 50 closed shell: Generalized seniority in cd, sn
  and te isotopes, Nuclear Physics A 992 (2019) 121619.

\bibitem{grodzins}
L.~Grodzins, The uniform behaviour of electric quadrupole transition
  probabilities from first 2$ sup+ $ states in even-even nuclei, Physics
  Letters 2 (1962).

\bibitem{pritychenko2017revisiting}
B.~Pritychenko, M.~Birch, B.~Singh, Revisiting grodzins systematics of b (e2)
  values, Nuclear Physics A 962 (2017) 73--102.

\bibitem{zuker2021quadrupole}
A.~Zuker, Quadrupole dominance in the light sn and in the cd isotopes, Physical
  Review C 103~(2) (2021) 024322.

\bibitem{lipkin1960collective}
H.~J. Lipkin, Collective motion in many-particle systems: Part 1. the violation
  of conservation laws, Annals of physics 9~(2) (1960) 272--291.

\bibitem{nogami1964improved}
Y.~Nogami, Improved superconductivity approximation for the pairing interaction
  in nuclei, Physical Review 134~(2B) (1964) B313.

\bibitem{suhonen2007nucleons}
J.~Suhonen, From nucleons to nucleus: concepts of microscopic nuclear theory,
  Springer Science \& Business Media, 2007.

\bibitem{maheshwari2016odd}
B.~Maheshwari, A.~K. Jain, Odd tensor electric transitions in high-spin
  sn-isomers and generalized seniority, Physics Letters B 753 (2016) 122--125.

\bibitem{hamamoto1965empirical}
I.~Hamamoto, Empirical data concerning the electric quadrupole moments, Nuclear
  Physics 73~(1) (1965) 225--233.

\bibitem{nair1977neutron}
S.~C.~K. Nair, A.~Ansari, L.~Satpathy, Neutron-proton interaction and nuclear
  deformations, Physics Letters B 71~(2) (1977) 257--258.

\bibitem{federman1977towards}
P.~Federman, S.~Pittel, Towards a unified microscopic description of nuclear
  deformation, Physics Letters B 69~(4) (1977) 385--388.

\bibitem{casten1985npnn}
R.~F. Casten, Npnn systematics in heavy nuclei, Nuclear Physics A 443~(1)
  (1985) 1--28.

\bibitem{heyde2011shape}
K.~Heyde, J.~L. Wood, Shape coexistence in atomic nuclei, Reviews of Modern
  Physics 83~(4) (2011) 1467--1521.

\bibitem{bonatsos2015proton}
D.~Bonatsos, I.~Assimakis, A.~Martinou, Proton-neutron pairs in heavy deformed
  nuclei, arXiv preprint arXiv:1510.01473 (2015).

\bibitem{lanealphawidhts}
A.~M. Lane, Nuclear theory, W. A. Benjamin Inc., New York, 1964.

\bibitem{flowers1963charge}
B.~Flowers, M.~Vuji{\v{c}}i{\'c}, Charge-independent pairing correlations,
  Nuclear Physics 49 (1963) 586--604.

\bibitem{mang1960calculation}
H.~J. Mang, Calculation of $\alpha$-transition probabilities, Physical Review
  119~(3) (1960) 1069.

\bibitem{sandulescu1962reduced}
A.~S\v{a}ndulescu, Reduced widths for favoured alpha transitions, Nuclear
  Physics 37 (1962) 332--343.

\bibitem{sandulescu2012four}
N.~Sandulescu, D.~Negrea, C.~W. Johnson, Four-nucleon $\alpha$-type
  correlations and proton-neutron pairing away from the {N= Z} line, Physical
  Review C 86~(4) (2012) 041302.

\bibitem{baran2016proton}
V.~V. Baran, D.~S. Delion, Proton-neutron versus $\alpha$-like correlations
  above sn 100, Physical Review C 94~(3) (2016) 034319.

\bibitem{delion2009universal}
D.~S. Delion, Universal decay rule for reduced widths, Physical Review C 80~(2)
  (2009) 024310.

\bibitem{delion2010investigations}
D.~S. Delion, R.~Wyss, R.~J. Liotta, B.~Cederwall, A.~Johnson, M.~Sandzelius,
  Investigations of proton-neutron correlations close to the drip line,
  Physical Review C 82~(2) (2010) 024307.

\end{thebibliography}

\end{document}